\pdfoutput=1
\documentclass[english,manuscript]{aastex}
\usepackage[T1]{fontenc}
\usepackage[latin9]{inputenc}
\usepackage{graphicx}

\makeatletter
\shorttitle{Low Luminosity AGN candidates in SDSS}
\shortauthors{Torres-Papaqui et al.}
\bibpunct[,]{(}{)}{;}{a}{,}{,}

\makeatother

\usepackage{babel}

\begin{document}

\title{Low Luminosity AGN candidates in SDSS}
\author{J.~P. Torres-Papaqui, R. Coziol, J.~M. Islas-Islas, R. A. Ortega-Minakata, and D. M. Neri-Larios}

\affil{Departamento de Astronom\'ia, Universidad de Guanajuato, Apartado Postal 144, 36000, Guanajuato, Gto., M\'exico (papaqui@astro.ugto.mx).}

% English abstract
\begin{abstract}
In a sample of 476931 NELGs obtained from the SDSS DR5
data we find that in 22\% of the galaxies the emission line
[OIII]$\lambda$5007, H$\beta$, or both, are missing. The nature of
the activity in these galaxies was determined using a diagnostic
diagram comparing the equivalent width of [NII]$\lambda$6584 with
the ratio [NII]$\lambda$6584/H$\alpha$. The majority of these
galaxies are AGN. The H$\alpha$ emission lines have a mean FWHM of
400 km s$^{-1}$ and mean luminosity of 5.6$\times$10$^{39}$ erg
s$^{-1}$, which justify their classification as LLAGN. A study of
their star formation histories using STARLIGHT reveals no trace of
star formation over the last Gyr period. The hosts of the LLAGNs are
early-type, T$\le 2$, with bulges more massive than those of the
luminous AGNs.
\end{abstract}

% Keywords must be from the standard list and in alphabetical order.
% You should have no more than SIX different keywords.

% Spanish abstract - leave blank and it will be translated by the
% editors.
\section{Resumen}
En una muestra de 476931 galaxias con l\'ineas de emisi\'on
agostas del Sloan Digital Sky Survey Data Release 5, identificamos y
estudiamos galaxias de las cuales algunas importantes l\'ineas de
emisi\'on usadas para determinar la naturaleza de su actividad
([OIII]$\lambda$5007\AA, H$\beta$, o ambas) no est\'an presentes.
Este fen\'omeno afecta al 22\% de las galaxias con l\'ineas de
emisi\'on y no esta relacionado con una baja raz\'on de se\~nal a
ruido. En el diagrama comparando la largura equivalente
EW([NII]$\lambda$6584) con la raz\'on de l\'inea
[NII]$\lambda$6584/H$\alpha$, la mayoria de estas galaxias se
clasifica como AGN. El FWHM de la l\'inea H$\alpha$ es del orden de
400 km s$^{-1}$ y la luminosidad mediana es 5.6$\times$10$^{39}$ erg
s$^{-1}$, lo cual justifica la clasificaci\'on de estas galaxias
como AGN de baja luminosidad. Un estudio en sus historias de
formaci\'on estelar usando STARLIGHT revela que no existe
formaci\'on estelar en el \'ultimo Giga a\~no. Las galaxias
anfitrionas de las LLAGNs son de tipo morfol\'ogico temprano con
bulbos m\'as masivos que los de AGN luminosas.

\section{Introduction and Discussion}
\label{sec:intro}

Many galaxies show narrow emission lines in their spectra. We call
them Narrow Emission Line Galaxies (NELGs). Following the seminal
studies of Baldwin, Phillips \& Terlevich (1981) and Veilleux \&
Osterbrock (1987) various spectroscopic diagnostic diagrams were
devised to determine the nature of their ionization sources. The
majority of NELGs traces a sequence in ionization consistent with OB
stars, which categorizes them as Star Forming Galaxies (SFGs). A
smaller fraction shows an excess of ionization that cannot be
explained by star forming activity. It is usually assumed that their
source of ionization is non thermal. These galaxies are generally
classified as Active Galactic Nuclei (AGNs). Galaxies falling in
between SFGs and AGNs are then classified as Transition Type Objects
(TOs).

In a sample of 476931 NELGs from the Sloan Digital Sky Survey Data
Release~5 (Adelman-McCarthy et al. 2007) we find three different
cases of NELGs where the most important emission lines used in
standard diagnostics diagrams are missing. We count 68491 galaxies
without H$\beta$, 27985 without [OIII] and 10926 without both lines.
This represents 22\% of the whole sample. Discarding NELGs with
S/N$<3$ (S/N$>$10 in the continuum) leaves 224846 galaxies (47\% of
the original sample). We now count 34307 galaxies without H$\beta$,
12455 without [OIII], and 2840 without both lines, again amounting
to 22\% of the sample. This shows that the absence of emission lines
is not a phenomenon related to low S/N.

\begin{figure}[!t]
  \includegraphics[width=0.8\columnwidth]{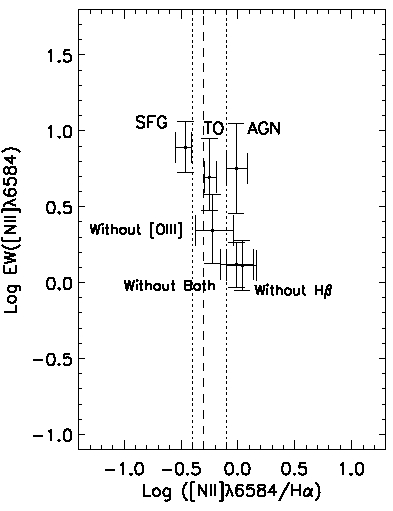}
  \caption{NII diagnostic diagram; shown are medians and quartiles of NELGs with emission lines missing, compared with NELGs classified using standard diagnostic diagrams.}
  \label{fig:01}
\end{figure}

Using the spectral synthesis code STARLIGHT (Cid Fernandes et al.
2005) the fluxes and EW in H$\alpha$ and [NII]$\lambda$6584 were
measured and compared to determine the activity type of the NELGs
with emission lines missing (Coziol et al. 1998). We summarize our
results in Figure~\ref{fig:01}. The majority of NELGs with emission
lines missing fall on the AGN side of the diagram. This is
particularly true for the galaxies without H$\beta$. The median
H$\alpha$ luminosity ranges between $1.7\times 10^{40}$ erg s$^{-1}$
for the galaxies without [OIII], and $5.6\times 10^{39}$ erg
s$^{-1}$ for the galaxies without H$\beta$. The low luminosity
justifies the classification of these galaxies as Low Luminosity
AGNs (LLAGNs).

The NELGs with emission lines missing have broader line profiles
than the SFGs. After correction for the resolution (Greene \& Ho
2006), the FWHM of H$\alpha$ fall between 25/75 percentiles 290 and
499 km s$^{-1}$. This is compared with 337 to 478 km s$^{-1}$ in
luminous AGNs and 253 to 292 km s$^{-1}$ in SFGs. The gas producing
the emission is moving at relatively higher speed in AGNs than in
SFGs.

The morphology of the galaxies were determined based on photometric
colors and inverse concentration index (Shimasaku et al. 2001).
While the galaxies without [OIII] are late-type spirals, T$\ge 3$,
the galaxies without H$\beta$ and with both lines missing are
dominantly early-type galaxies, T$\le2$. In general, the fraction of
early type galaxies increases rapidly as the EW decreases.
Consistent with the morphology classification, the star formation
histories as determined by STARLIGHT indicate that the NELGs without
H$\beta$ and those with both lines missing show no evidence of star
formation activity since $t=10^9$yrs.

\begin{figure}[!t]
  \includegraphics[width=0.8\columnwidth]{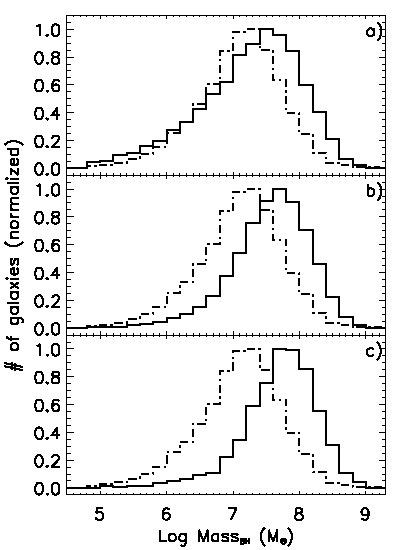}
  \caption{Distributions of the bulge masses:  a) galaxies without [OIII]; b) galaxies without H$\beta$; c) galaxies without both lines. The point dashed line corresponds to the luminous AGNs.}
  \label{fig:02}
\end{figure}

We compare the mass of the bulges as found in the three different
samples of NELGs with emission lines missing in Figure~\ref{fig:02}.
The bulges in the galaxies without H$\beta$ and with both lines
missing are generally more massive than in the luminous AGNs.

In their study of the activity of compact groups of galaxies Coziol
et al. (1998) and Mart\'{\i}nez et al. (2008; 2010) have found many
examples of NELGs with emission lines missing, consistent with
LLAGNs. Previous observations by Phillips et al. (1986) have also
shown this phenomenon to be common in clusters of galaxies,
affecting between 55\% to 60\% of all early-type galaxies. Although
we do not have a definite evidence yet, our preliminary analysis
suggests that more than 50\% (87\% for the galaxies without both
lines) are in dense galactic environments. Therefore, the LLAGNs
phenomenon may be related to a particular evolutionary phase of
galaxies that formed in dense galactic environments.

There is presently no consensus among researchers on what is the
physical nature of AGNs with low luminosity. However, according to
the standard interpretation, an AGN is produced by the accretion of
matter onto a massive black hole. Therefore, one possibility to
explain the LLAGNs would be that the strong astration rates
necessary to produce the massive bulges of galaxies in dense
galactic environments have depleted the galaxies of most of their
gas, affecting the accretion rates of their black holes.

We also acknowledge support from PROMEP (Grant No. 103.5-10-4684).

\end{document}